\documentclass[12pt]{article}
\usepackage{xcolor}
\usepackage{amsfonts}
\usepackage{amssymb}
\usepackage{amscd}
\usepackage{amsmath}
\usepackage{epsfig}
\usepackage{graphicx, subfigure}
\usepackage{latexsym}
\usepackage{pst-3dplot}
\newlength{\dinwidth}
\newlength{\dinmargin}
\newtheorem{theorem}{Theorem}

\newtheorem{corollary}{Corollary}
\newtheorem{remark}{Remark}
\newtheorem{lemma}{Lemma}
\newtheorem{example}{Example}

\def \i{{\rm i}}

\begin{document}

\title{Topological recursion of Eynard-Orantin and the Harmonic Oscillator}
\author{Miguel Cutimanco$^1$, Patrick Labelle$^2$ and Vasilisa Shramchenko$^1$}

\date{}

\maketitle

\footnotetext[1]{Department of mathematics, University of
Sherbrooke, 2500, boul. de l'Universit\'e,  J1K 2R1 Sherbrooke, Quebec, Canada. E-mail: {\tt Miguel.Cutimanco.Panduro@usherbrooke.ca ; Vasilisa.Shramchenko@Usherbrooke.ca}}

\footnotetext[2]{Champlain Regional College, Lennoxville campus, Sherbrooke, Quebec, Canada. E-mail: {\tt plabelle@crc-lennox.qc.ca}}

\begin{abstract}

We apply the Chekhov-Eynard-Orantin topological recursion to the curve corresponding to the quantum harmonic oscillator and demonstrate that the result is equivalent to the WKB wave function. We also show that using the multi-differentials obtained by the topological recursion from the harmonic oscillator curve, one generates naturally the so-called Poincar\'e polynomials associated with the orbifolds of the metric ribbon graphs.

\end{abstract}

%\tableofcontents

%\newpage

\section{Introduction}

The topological recursion of Eynard-Orantin (also called Chekhov-Eynard-Orantin recursion)  defined in \cite{EO} is a way to associate an infinite hierarchy of interesting multivariable differentials $w_{g,n}$ to an algebraic curve. This recursion appeared in the context of matrix models where the algebraic curve in question was the spectral curve of the model, see for example \cite{Chekhov2,Chekhov3}. In \cite{EO} this procedure is described for an arbitrary algebraic curve, not necessarily related to any matrix model, and it turned out that for suitably chosen curves the Eynard-Orantin recursion formula coincides with (often Laplace transformed) known recursions, such as Mirzakhani recursion \cite{Mirzakhani} for Weil-Petersson volumes of the moduli space $\mathcal M_{g,n}$ of curves with $n$ marked points, the cut-and-join recursion of \cite{GouldenJackson} on Hurwitz numbers, see \cite{Borot,BouchardMarino, DB, Do, EMS, Norbury1}, Gromov-Witten invariants of $\mathbb P^1$, see \cite{DB1, DB2, FLZ, Norbury3}, and many others, see \cite{DumiMulasereview, EynardICM,EOJPA,Mulase,Norburyreview} for  overviews. The topological recursion on the curve given by $y^2=x$ recovers the DVV recursion \cite{DVV} for the intersection numbers of $\psi$-classes of the Deligne-Mumford compactification of the $\mathcal M_{g,n}$, see  \cite{BCSW} and \cite{Zhou1}.

Starting with an algebraic curve given by $F(x,y)=0$ such that $x$ is a degree two function on the curve 
one can quantize it, that is replace the coordinates $x,y$ of the {\it classical} curve by operators $\hat{x}, \hat{y}$ thus obtaining a {\it quantum} curve. With the  choice of quantization prescription $\hat{x} = x $ and $\hat{y} = \hbar \partial/\partial x$
 the curve becomes  a second order differential operator which can be applied to a complex wave function to produce a  Schr\"odinger-like equation. Not all quantized curves correspond to physically meaningful systems, however, since  the resulting operators must be Hermitian (in the space of square integrable complex wave functions) in order to correspond to {\it bona fide} Hamiltonians. 

It has been observed  that  for many mathematically interesting examples
(\cite{BergereEynard, DB2, DumitrescuMulase, GukovSulkowski, MulaseSulkowski, Zhou2}), the topological recursion applied to the classical curve can be used to generate the WKB expansion \cite{EB, Brillouin, Dunham, Jeffreys, Kramers, Wentzel} of the  wave function. Again, strictly speaking this corresponds to a physically meaningful solution only for Hermitian quantum curves.

Motivated by the connection of the topological recursion with Schr\"odinger equation, we consider a particularly important system, that of the harmonic oscillator.
From the point of view of physics, the  importance of the harmonic oscillator comes from the fact that it is  a nontrivial exactly soluble system as well as a physically meaningful one  since a quadratic potential is a good approximation to many phenomenological potentials in atomic, molecular and nuclear physics. But its importance goes far beyond this as most perturbative quantum field theory calculations are based on expansions around noninteracting quantum fields described by infinite numbers of harmonic oscillators. As such, the harmonic oscillator is a cornerstone of both quantum field theory and particle physics.

 The Schr\"odinger equation in this case corresponds to the curve $y^2=x^2-c^2$ with $c^2$  proportional to  the energy of the system. We call this curve the {\it harmonic oscillator curve}. 
 With the above choice of quantization prescription, the quantum curve leads to the Schr\"odinger equation for the quantum harmonic oscillator.

Our first result is the proof of the fact that the WKB approximation for the wave function of the harmonic oscillator can be obtained from the Eynard-Orantin recursion formalism applied to the harmonic oscillator curve.  We are grateful to V. Bouchard for pointing to us that this result has been recently demonstrated in \cite{EB} using a different approach and for a large class of other curves.

Moreover, it turns out that the Eynard-Orantin topological recursion on the harmonic oscillator curve also corresponds to a recursion known in a seemingly different context, namely, it corresponds to the Laplace transformed recursion for the number $N_{g,n}(p_1,\dots,p_n)$ of ribbon graphs of genus $g$ with $n$ faces such that a given positive integer $p_i$ is assigned to $i$th face. The recursion on $N_{g,n}$ was derived in \cite{Chapman}. Moreover, by Laplace transforming  $N_{g,n}$ in \cite{Chapman} the functions $L_{g,n}(\omega_1,\dots, \omega_n)$ of complex variables $\omega_i$ are introduced. It is shown in \cite{MulasePenkava} that after a suitable change of variables $(\omega_1,\dots, \omega_n)\mapsto {\bf z}= (z_1,\dots, z_n)$, the functions $L_{g,n}(\omega_1({\bf{z}}), \dots, \omega_n({\bf{z}}) )$ for the pair $(g,n)$ satisfying the stability condition $2g-2+n>0$ coincide with a polynomial $P_{g,n}(z_1,\dots, z_n)$ associated with the orbifold $RG_{g,n}$ of metric ribbon graphs of type $(g,n)$. This polynomial was called in \cite{MulasePenkava} the {\it Poincar\'e polynomial} of $RG_{g,n}$.

We have found  that the differentials $w^H_{g,n}$ obtained by topological recursion of Eynard-Orantin for the harmonic oscillator curve 
are also closely linked to  invariants of the moduli spaces $\mathcal M_{g,n}$ and spaces of metric ribbon graphs $RG_{g,n}$. Namely, the $w^H_{g,n}$ are Laurent polynomials symmetric in all the variables and invariant under the simultaneous map $z_i\mapsto -\frac{1}{z_i}$ for all $i=1,\dots, n$. The top degree homogeneous part of $w^H_{g,n}$ coincides, up to an overall factor containing $c^2$, with the corresponding $w_{g,n}$ obtained for the Airy curve, which implies that the coefficients of the top degree homogeneous part of $w^H_{g,n}$ contain  the intersection  numbers $\langle  \tau_{a_1},\dots, \tau_{a_n} \rangle$ of $\psi$-classes of the Deligne-Mumford compactification $\overline{\mathcal M}_{g,n}$ of the moduli space of stable curves, see Section \ref{sect_HO}. Moreover, we find that  the Poincar\'e polynomial of $RG_{g,n}$ from \cite{MulasePenkava} for $2g-2+n>0$ is proportional to the integral of the $w^H_{g,n}$ over all the variables and that a certain choice of value for $c^2$ can be made to set the constant of proportionality equal to one, giving 
\begin{equation}
\label{ints}
 P_{g,n}(z_1,\dots, z_n) = \int_{-1}^{z_1}\dots \int_{-1}^{z_n} w^H_{g,n}(z_1,\dots, z_n).
\end{equation}

It must be noted that the differentials 
\begin{equation}
\label{intro_P}
d_{z_1}\dots d_{z_n}P_{g,n}(z_1,\dots, z_n) =w^H_{g,n}(z_1,\dots, z_n)
\end{equation}
 for $2g-2+n\geq 2$ were obtained  in \cite{Chapman} by a modified Eynard-Orantin recursion on the curve $xy=y^2+1$; formula \eqref{ints} however seems to be new.  The recursion from \cite{Chapman} is equivalent to our recursion on the curve $y^2=x^2-c^2$ with $c^2=2$ up to two important comments. First, the Eynard-Orantin recursion starts with a curve and two initial differentials, $w_{0,1}(P)=ydx$ and $w_{0,2}(P_1,P_2)$ where $P, P_1, P_2$ are points on the curve. The differential $w_{0,2}$ as defined originally in \cite{EO} is a symmetric bidifferential with a second order pole at $P_1=P_2$ of vanishing residue and no other singularities. This definition has to be modified for the curve of \cite{Chapman} in order to obtain the differentials \eqref{intro_P}
 whereas the recursion on our harmonic oscillator curve produces these differentials according to the original definition of Eynard-Orantin. Moreover, in \cite{Chapman}, more initial data has to be provided for the recursion, namely, together with $w_{0,1}$ and $w_{0,2}$, also the differentials $w_{1,1}$ and $w_{0,3}$, that is those with $2g-2+n=1$ have to be given. For the harmonic oscillator curve however, starting with the initial data of $w_{0,1}$ and $w_{0,2}$,  the Eynard-Orantin recursion computes $w^H_{1,1}$ and $w^H_{0,3}$ as well as all the other $w^H_{g,n}$ with $2g-2+n>0$ such that \eqref{intro_P} holds. We believe that because of these reasons, the harmonic oscillator curve is the more appropriate curve to look at in this context.

The paper is organized as follows. In Section \ref{sect_recursion} we briefly describe the simplest case of  Eynard-Orantin definition of the topological recursion procedure, the case of hyperelliptic curves. In Section \ref{sect_WKB} we describe the WKB method for finding wave functions and energies of a quantum mechanical system. In Section \ref{sect_Airy} we present a summary of the topological recursion for the so-called Airy curve. In Section \ref{sect_HO} we introduce the harmonic oscillator curve, the initial data for the topological recursion on this curve and state the theorem relating the WKB analysis for the wave function of the harmonic oscillator and the differentials $w^H_{g,n}$ obtained by the Eynard-Orantin recursion. We also discuss the energy quantization for harmonic oscillator from the  point of view of the curve and give a general form of the $w^H_{g,n}$ exhibiting the relationship between these differentials and the corresponding ones computed by the recursion on the Airy curve. This shows that some coefficients in $w^H_{g,n}$ are given by the intersection numbers of $\psi$-classes on $\overline{\mathcal M}_{g,n}$.  In Section \ref{sect_proof} we prove the theorem stated in the previous section. In Section \ref{sect_ribbon} we describe links between the topological recursion on the harmonic oscillator curve and orbifolds of the metric ribbon graphs. In the Appendix we prove a technical lemma from Section \ref{sect_proof}.

\vskip 1cm

{\bf Acknowledgements.}  The authors are grateful to Krishna Kappagantula and  Motohico Mulase for many useful discussions and explanations of various concepts and to Vincent Bouchard for pointing out reference \cite{EB}.  P.L. is grateful to  the Fonds  de recherche du Qu\'ebec - Nature et technologies (FQRNT)  for a grant as part of the Programme de recherches pour les chercheurs de coll\`ege and to the STAR research cluster of  Bishop's University.   V.S. gratefully acknowledges
support from the Natural Sciences and Engineering Research Council of Canada as well as the University of Sherbrooke. P.L. and V.S.  thank the  Max Planck Institute for Mathematics in Bonn, where part of this work was done, for hospitality and a perfect working environment.

\section{Topological recursion of Eynard-Orantin}
\label{sect_recursion}

The Eynard-Orantin topological recursion is a procedure that produces a hierarchy of multi-differentials defined on a given algebraic curve. We briefly describe this algorithm in the simplest case of a hyperelliptic curve defined by a polynomial of the form 
\begin{equation*}
y^2=\prod_{j=1}^m (x-x_j).
\end{equation*}
Let us denote by $\mathcal L$ the Riemann surface of genus $g$ associated with this algebraic curve. We thus have two meromorphic functions $x$ and $y$ on $\mathcal L$. The ramified covering of $\mathbb {CP}^1$ defined by the function $x:\mathcal L \to \mathbb{CP}^1$ has simple ramification points $A_i\in\mathcal L$, the points corresponding to $x=x_i$ and $y=0$. We denote by $q^*=(x(q),-y(q))$ the hyperelliptic involution of the point $q=(x(q),y(q))$. 

Given a symplectic basis of homology cycles $\{a_i, b_i\}_{i=1}^g$,
the topological recursion algorithm defines an infinite family of differentials $w_{g,n}(p_1, \dots, p_n)$ with $p_i\in\mathcal L$ and integers $g\geq 0$, $n\geq 1$, where $w_{g,n}$ is a differential in each of the $p_i$. The initial data for the recursion consists of the differential $w_{0,1}$ and the bidifferential $w_{0,2}$ defined as follows.

\begin{itemize}
\item $w_{0,1}(p)=ydx$ where $p$ corresponds to the point $(x,y)$ on the algebraic curve. 
\item $w_{0,2}(p,q)$  is a symmetric bidifferential with the double pole at $p=q$ of the local form 
	\begin{equation*}
	 w_{0,2}(p,q)=\Big{(}\left(\tau(p)-\tau(q)\right)^{-2} + \mathcal O(1)\Big{)}d\tau(p)d\tau(q),
	\end{equation*}
at $p\sim q$ in terms of  a local parameter $\tau$. 
\end{itemize}

Starting with these data, we define the recursion kernel $K(q,p)$ for  $p,q\in\mathcal L$,
\begin{equation}
\label{K}
K(q,p)=\frac{1}{2}\frac{\int_q^{q^*}w_{0,2}(\xi,p)}{(y(q)-y(q^*) )dx(q)},  
\end{equation}
and the differentials $w_{g,n}$ for $(g,n)\notin\{(0,1),(0,2)\}$, that is for $2g-2+n>0$:
\begin{multline}
\label{recursion}
w_{g,n}(p_1,\dots, p_n) =
\\
= \sum_{i=1}^m \underset{q=A_i}{\rm res} K(q,p_1)\left(  w_{g-1,n+1}(q,q^*, p_2, \dots, p_n) 
+ \sum_{\substack{g_1+g_2=g\\ I\coprod J =\{2,\dots, n\}}}^{{\rm no }\;(0,1)} w_{g_1,|I|+1}(q,p_I)w_{g_2,|J|+1}(q^*,p_J)
\right).
\end{multline}
Here $I$ {\small${\coprod}$} $J$ is the disjoint union, $|I|$ denotes the number of elements in the set $I$ 
and the second sum  excludes the value $(0,1)$  for $(g_1,k)$ and $(g_2,k)$.
This is a recursion with respect to the number $2g-2+n>0.$
The obtained multi-differentials $w_{g,n}$ are invariant under arbitrary permutation of their arguments $z_1,\dots, z_n$, have poles at the ramification points $A_i$ with respect to each of the arguments and no other singularities, see \cite{EO}.

\section{WKB method}
\label{sect_WKB}

The WKB method is a procedure for finding approximate solutions to differential equations containing a small parameter multiplying the 
highest order derivative. A special case of such an equation is the one-dimensional time independent Schr\"odinger equation,
\begin{equation*}
- \frac{\hbar^2}{2M} \frac{d^2 \Psi(x,\hbar)}{dx^2} + V(x) \Psi(x,\hbar) = E \,  \Psi(x, \hbar )  \, ,\end{equation*}
where $V(x)$ is the potential energy, $M$  the mass of the particle and $E$  its energy.
Here Planck's constant $\hbar$ plays the role of the small parameter. Let us define 
\begin{equation*} f(x) = 2  M V(x) 
\end{equation*}
and

\begin{equation*}
 c^2 = 2  M  E \, ,
\end{equation*}
so that Schr\"odinger's equation becomes
\begin{equation*}
\hbar^2  \,  \frac{d^2 }{dx^2}   \Psi(x,\hbar) =  \left( f(x) - c^2  \right) \, \Psi (x,\hbar) .
\end{equation*}
The equation can be solved exactly only for a few potentials. 
This paper will discuss two of the simplest cases, the Airy potential, for which $f(x) = \kappa \, M \, x$, and the harmonic oscillator potential, for which  $f(x) = M^2\, \omega^2 x^2 $, where $\kappa$ and $\omega$ are constants having dimensions of energy per unit length and inverse time, respectively. In the following we will set  the products  $\kappa \, M$  and    $M \omega$ equal to one;  it is easy to reinstate them using dimensional analysis if needed.

 A physical wave function must be square integrable, namely $\int_{-\infty}^\infty  |\Psi(x)|^2 dx  $ must be  finite. For confining potentials, {\it i.e.} potentials which obey $V(\pm \infty) > c^2$,  as is the case for the harmonic oscillator, it is possible to obtain square integrable solutions only for discrete values of $c^2$; this is the origin of energy quantization.

The Airy potential is not confining so the energy is not quantized and we are free to set the value of $c^2$ to whatever value we 
desire. In the following we will set $c^2=0$ for the Airy potential.  No solution of the Airy Schr\"odinger equation is square integrable but it is possible to construct linear combinations of solutions (corresponding to different values of $c^2$) that are normalizable at the condition of considering  for solution only the Airy function $Ai(x)$ and rejecting the so-called ``Bairy" function $Bi(x)$ which diverges as $x$ goes to $+ \infty$.

 The square integrable solutions of the harmonic oscillator Schr\"odinger equation are given by  exponentials times the Hermite polynomials, more precisely:
\begin{equation}
\Psi_n(x,\hbar)  = \exp \left(- \frac{ x^2}{2 \hbar} \right)  ~ H_n( \sqrt{\hbar}  x )  \, , \label{herm}
\end{equation}
where 
the overall constant has been chosen such that    $\int_{-\infty}^\infty  |\Psi_n(x)|^2 dx = 1$ and $n=0,1,2 \ldots$ is referred to as a quantum number. These solutions correspond to 
quantized energies according to 
\begin{equation}
E_n (\hbar) = \left( n+ \frac{1}{2} \right)  \, \hbar \omega  \, .
\label{eho} \end{equation}

For most potentials, exact solutions are not available and one must rely on approximation methods. The  WKB method  (developed by Wentzel \cite{Wentzel}, Kramers \cite{Kramers},  Brillouin \cite{Brillouin} and, independently, by the mathematician H. Jeffreys \cite{Jeffreys}) consists in using the following ansatz for the wave function:
\begin{equation}
\label{WKB}
\Psi(x,\hbar) =  {\rm exp} \sum_{m=0}^\infty \hbar^{m-1}S_m(x) \, ,
\end{equation}
which defines the functions $S_m(x)$. These are obtained by using the ansatz in Schr\"odinger's equation and imposing 
that the two sides agree order by order in $\hbar$.  This provides an  expansion of the wave function away from the so-called ``classical turning points", defined as the points  with coordinates $x_t$ such that   $f(x_t) = c^2$.
  The terminology comes from the fact that these are the points where a classical particle moving
in the potential 
would change direction.  For the Airy potential (with $c^2=0$), there is only one classical turning point, $x_t =0$ and  for the harmonic oscillator, there are two turning points at $x_t = \pm c$.

To lowest order in $\hbar$, one obtains

\begin{equation}
S_0' (x) =  \pm \sqrt{f(x) - c^2 }   \, \label{tworoots}
\end{equation}
where the prime denotes a derivative with respect to $x$. Therefore
\begin{equation*}
S_0(x) = \pm  \int^x   \,  \sqrt{f(\tilde{x}) - c^2 }  \,d\tilde{x}\,.  \end{equation*}
 In carrying out the integration, any constant of integration can be ignored since the  overall normalization of the wave function is not determined by the WKB expansion.

Once $S_0'(x)$ is determined, all the $S_m'(x)$ are fixed. Indeed, by equating terms of the same order in $\hbar$ in  Schr\"odinger's equation, we get (suppressing the $x$ dependence of the $S'_m$) 
\begin{equation}
S_{m+1}' = -\frac{\left(S_m'' +  \sum_{i=1}^{m} S_i' S_{m+1-i}' \right) }{2 S_0'}  \, ,\label{wkb1}
\end{equation}
where $m \geq 0$.
In particular, consider $S_1'$ which will play a special role in the context of topological recursion. It is given by
\begin{equation*}
S_1' = - \frac{S_0''}{2 S_0'} = - \frac{1}{2} \frac{d \ln  S_0' }{dx} \, .
\end{equation*}
It will prove convenient later to isolate the contributions of $S_0'$ and $S_1'$ in  (\ref{wkb1})  and  give the recursion formula for the $S_m'$  with $m \geq 2$:

\begin{equation}
\label{WKB_WKB}
S_{m}' =- \frac{S_{m-1}''}{2S_0'}   - \frac{S_1' S_{m-1}'}{S_0'}   + \delta_{m,2}  \frac{(S_1')^2}{2S_0'}
-\sum_{\substack{i+j=m\\i,j\geq 2}}   \frac{S_i' S_{j}'}{2 S_0'} \, .
\end{equation}

We have  discussed how the wave functions are calculated  using the WKB method, but so far  the value of $c^2$ has been left completely arbitrary and hence the formalism  does not say anything about the allowed values of energy. This information comes from a generalization of the so-called Bohr-Sommerfeld quantization condition. For confining potentials for which there are only two classical turning points, the energies can be determined from the WKB solution by imposing \cite{Dunham}
\begin{equation}
\frac{1}{2 \pi i} \oint \sum_{m=0}^\infty  \hbar^{m-1} S_m'(x) \, dx  = n \, , \label{quant}
\end{equation} 
where $n$ is an integer and 
 the contour of integration encircles the two  turning points counterclockwise (when the potential has discontinuities, this formula must be corrected but this will not be relevant to the present paper).  This leads to an expansion for $E_n(\hbar)$ in powers of $\hbar$.  It may be shown that the $m=1$ contribution to the left hand side of  (\ref{quant}) gives $-1/2$ while  the $S_m'$ for odd  $m>1$ are  total derivatives of  single valued functions so  they do not contribute to the integral. We are then left with

\begin{equation}
\frac{1}{2 \pi i} \oint \sum_{m=0}^\infty  \hbar^{2m-1} S_{2m}'(x)  \,  dx  = n + \frac{1}{2} \, .
\label{energy}
\end{equation} 

\begin{example}
\label{example_Sk}
To be concrete, let's consider the harmonic oscillator in more  details.  The classical turning points are then at $x_t = \pm c $.  Consider the region $ x > c$, in which case we must choose the negative sign in \eqref{tworoots} to obtain square integrable solutions.  A straightforward calculation gives 
\begin{eqnarray*} 
S_0(x) &=&   \frac{c^2}{2} \ln \bigl(  \sqrt{x^2-c^2}  + x \bigr)-\frac{x}{2} \sqrt{x^2-c^2}  ,
\\ S_1(x) &=&   - \frac{1}{2}  \ln \bigl(\sqrt{x^2-c^2} \bigr) \;,
\\ S_2(x) &=& \frac{x^3-6c^2x}{24 c^2 (x^2-c^2)^{3/2}}\;,
\\ S_3(x) &=& \frac{3x^2+2c^2}{16(x^2-c^2)^3}
\end{eqnarray*}
and so on, up to the addition of constants which are not determined by the WKB method.  Using these expressions into the WKB wave function (\ref{WKB}), one obtains an expansion of the exact wave function in powers of $\hbar$. Note that by dimensional analysis, $c^2$ must be proportional to $\hbar$ so that the classical turning points are of order  $ x_t \simeq \pm \sqrt{\hbar} $ (or $\pm \sqrt{\hbar} /  M \omega $ if we reinstate these constants), therefore consistency requires that in addition to the explicit expansion in powers of $\hbar$ appearing in (\ref{WKB}), one must also perform an expansion in powers of $c^2/x \simeq \hbar/x $ to the same order. One then finds that the WKB expansion indeed agrees with the exact solution 
(\ref{herm})  expanded to the same order.

Consider now the calculation of the energies. The 
integral on the left hand side of (\ref{energy}) gives $c^2/2 \hbar $ for $m=0$ and zero for $ m>0$ therefore reproducing (\ref{eho}) after 
$\omega$ has been reinstated using dimensional analysis.
\end{example}

\section{Airy functions}
\label{sect_Airy}

A very interesting example for the topological recursion procedure is that of the Airy spectral curve
\begin{equation}
\label{Airy_curve}
y^2=x.
\end{equation}
Its quantization $\hat{x}=x$, $\hat{y}=\hbar \frac{\partial}{\partial x}$ gives the Airy differential equation
\begin{equation*}
\hbar^2\frac{\partial^2}{\partial x^2}\Psi(x,\hbar)=x\Psi(x, \hbar) \, ,
\end{equation*}
which corresponds to the Schr\"odinger equation with the Airy potential and $E=0$.

It was shown in \cite{BergereEynard} and \cite{Zhou2} that the topological recursion \eqref{recursion} for the Airy curve reproduces the WKB approximation for the solution of the Airy equation in the following sense. 

A solution to the Airy equation is represented by the expansion of the form \eqref{WKB}
and the coefficients $S_m$, starting from $m=2$, are obtained using the topological recursion on the Airy curve as follows.
The topological recursion for the Airy curve computes the differentials $w_{g,n}$ starting from the initial data
\begin{equation}
\label{Airy_initial}
w_{0,1}(p_1) = y_1dx_1, \qquad w_{0,2}(p_1, p_2) = \frac{dy_1 dy_2}{(y_1-y_2)^2},
\end{equation}
where $y$ is the local parameter away from the point at infinity of the Airy curve and $p_i$ is the point of the curve corresponding to $x=x_i$ and $y= y_i$. The {\it free energies} $F_{g,n}$ for the pairs $(g,n)$ in the stable range, that is $2g-2+n>0$, are defined as the following $n$-fold integrals of the differentials $w_{g,n}$:
\begin{equation}
\label{Fgn}
F_{g,n}(y_1,\dots, y_n) =\frac{1}{2^n} \int^{y_1}_{-y_1} \dots \int^{y_n}_{-y_n} w_{g,n}.
\end{equation}
Then the coefficients $S_m$, choosing the branch $y=\sqrt{x}$, are given by
\begin{eqnarray}
&&S_0(x) = -\frac{1}{2}\int^y_{-y} y' \, dx' = -\frac{2}{3}x^{\frac{3}{2}}\,;
\nonumber
\\
\nonumber
\\
&&S_1(x) = -\frac{1}{2}{\rm log}\,y  = -\frac{1}{4}{\rm log}\,x\,;
\nonumber
\\
\nonumber
\\
&&S_m(x) = \sum_{2g+n-2=m-1} \frac{1}{n!} F_{g,n}(y,\dots,y)\,, \qquad m\geq 2.
\label{Sm}
\end{eqnarray}
A simple proof of this fact for the Airy curve given in \cite{Zhou2} is based on the fact that the differentials $w_{g,n}$ with $2g-2+n>0$ for the Airy curve have a simple form
\begin{equation*}
w_{g,n}(z_1,\dots, z_n) =\sum_{\substack{a_1,\dots, a_n\geq 0\\\sum_{i=1}^na_i=3g+n-3}} c_{a_1,\dots, a_n}^g\prod \frac{1}{z_i^{2a_i+2}}.
\end{equation*}

Moreover, there is the remarkable observation from  \cite{BCSW}, \cite{EOJPA}  and \cite{Zhou1} that the coefficients $c_{a_1,\dots, a_n}^g$ contain information about the Deligne-Mumford compactification $\overline{\mathcal M}_{g,n}$ of the moduli space of stable curves. More precisely, the form of the differentials $w_{g,n}$ produced by the topological recursion of Eynard-Orantin for the Airy curve is
\begin{equation*}
w_{g,n}(y_1,\dots, y_n) = \frac{1}{2^{2g-2+n}} \sum_{\substack{a_1,\dots, a_n\geq 0\\\sum_{i=1}^na_i=3g+n-3}} \langle\tau_{a_1},\dots,\tau_{a_n}\rangle_g\prod_{i=1}^n \frac{(2a_i+1)!!}{y_i^{2a_i+2}},
\end{equation*}
where $\langle\tau_{a_1},\dots,\tau_{a_n}\rangle_g$ is the intersection number of the $\psi$-classes of $\overline{\mathcal M}_{g,n}$ and $2g-2+n>0$. As a consequence, the functions $F_{g,n}$ for the indices in the stable range have the form
\begin{equation}
\label{FgnAiry}
F_{g,n}(y_1,\dots, y_n) = \frac{(-1)^n}{2^{2g-2+n}} \sum_{\substack{a_1,\dots, a_n\geq 0\\\sum_{i=1}^na_i=3g+n-3}} \langle\tau_{a_1},\dots,\tau_{a_n}\rangle_g\prod_{i=1}^n \frac{(2a_i-1)!!}{y_i^{2a_i+1}}.
\end{equation}
It was shown in \cite{MulasePenkava} that these functions $F_{g,n}$ coincide up to an overall factor with the leading homogeneous part of a certain Laurent polynomial associated with the 
space of metric ribbon graphs $RG_{g,n}$, the so-called Poincar\'e polynomial of $RG_{g,n}$. We discuss this connection in more detail in Section \ref{sect_ribbon} for the functions $F^H_{g,n}$ the top homogeneous parts of which coincide with $F_{g,n}$ \eqref{FgnAiry}. 
\vskip 0.5cm
\begin{example} 
\label{example_Airy}
{\rm
We give a few examples of the differentials $w_{g,n}$ computed for the Airy curve \eqref{Airy_curve} with initial data \eqref{Airy_initial} for the recursion. We work in terms of  $y$ which is a local parameter away from the point at infinity $(x=\infty,y=\infty)$ of the curve.  The recursion kernel   \eqref{K} is then
\begin{equation*}
K^{Airy}(y,y_1) = \frac{1}{4}\frac{dy_1}{(y^2-y_1^2)ydy}
\end{equation*}
and we obtain the following multi-differentials. Denote $dy_{[k]} = dy_1\dots dy_k.\;$ For $2g-2+n=1$ we have
\begin{eqnarray*}
&&w_{0,3}(y_1,y_2,y_3)= \frac{dy_{[3]}}{2\,y_1^2y_2^2y_3^2}\,,
\\
&& w_{1,1}(y_1)= \frac{dy_1}{16\, y_1^4}.
\end{eqnarray*}
For $2g-2+n=2$ we have
\begin{eqnarray*}
&& w_{0,4} (y_1,y_2,y_3,y_4)= \frac{3\,dy_{[4]}}{4\, y_1^2y_2^2y_3^2y_4^2} \sum_{i=1}^4\frac{1}{y_i^2}\,,
\\
&& w_{1,2}(y_1,y_2) =\left( \frac{5}{32\, y_1^2y_2^6} +\frac{3}{32\, y_1^4 y_2^4} + \frac{5}{32\, y_1^6y_2^2}\right) dy_{[2]}.
\end{eqnarray*}
For $2g-2+n=3$ we have
\begin{eqnarray*}
&& w_{0,5}(y_1,y_2,y_3,y_4,y_5) = \frac{3\,dy_{[5]}}{2^3\,y_1^2y_2^2y_3^2y_4^2y_5^2} \left( 5\sum_{i=1}^5\frac{1}{y_i^4} + 6\sum_{i\neq j} \frac{1}{y_i^2y_j^2} \right)\,,
\\
&& w_{1,3}(y_1,y_2,y_3) = \frac{dy_{[3]}}{2^5\, y_1^2y_2^2y_3^2} \left( \frac{35}{2}\sum_{i=1}^3\frac{1}{y_i^6} + 15\sum_{i\neq j} \frac{1}{y_i^2 y_j^4} + \frac{9}{y_1^2y_2^2y_3^2}\right)\,,
\\
&&w_{2,1}(y_1) = \frac{105\,dy_1}{2^{10} y_1^{10}}.
\end{eqnarray*}
The integration \eqref{Fgn} of these differentials gives the functions $F_{g,n}\,$, for example
\begin{eqnarray*}
&&F_{0,3}(y_1,y_2,y_3)= -\frac{1}{2\,y_1y_2y_3};
\qquad\qquad\qquad\quad\;
 F_{1,1}(y_1)= -\frac{1}{3}\frac{1}{16\, y_1^3};
\\
&& F_{0,4} (y_1,y_2,y_3,y_4)= \frac{1}{4\, y_1y_2y_3y_4} \sum_{i=1}^4\frac{1}{y_i^2};
\qquad\quad
  F_{1,2}(y_1,y_2) = \frac{1}{32\, y_1y_2^5} +\frac{1}{3}\frac{1}{32\, y_1^3 y_2^3} + \frac{1}{32\, y_1^5y_2}.
\end{eqnarray*}
The principal specialization $F_{g,n}=F_{g,n}(y,\dots, y)$ gives us the coefficients $S_m$ for $m\geq 2$ according to \eqref{Sm}: 
\begin{eqnarray*}
&& S_2(x)= \frac{1}{3!}F_{0,3}+F_{1,1}
=-\frac{1}{12\, y^3} -\frac{1}{48\,y^3}
=-\frac{5}{48\,y^3}=-\frac{5}{48\,x^{3/2}};
\\
&&S_3(x)=\frac{1}{4!}F_{0,4} + \frac{1}{2}F_{1,2}
=\frac{1}{4!}\frac{1}{y^6} + \frac{1}{2} \frac{7}{3\cdot 32}\frac{1}{y^6}
= \frac{5}{64\,y^6}= \frac{5}{64\,x^3};
\\
&&S_4(x)=\frac{1}{5!}F_{0,5} + \frac{1}{3!}F_{1,3}+F_{2,1}
=-\frac{1}{5!}\frac{35}{2^3\,y^9} - \frac{1}{3!} \frac{83}{2^6\cdot 3 \,y^9} - \frac{5\cdot 7}{2^{10}\cdot 3\,y^9}
=-\frac{1105}{2^{10}\cdot3^2\,x^{9/2}}\,.
\end{eqnarray*}
With these coefficients and the $S_0$, $S_1$ from \eqref{Sm}, the WKB approximation formula \eqref{WKB} gives, up to an overall factor,  the asymptotic expansion of the Airy function ${\rm Ai}(x)$ for large positive $x$. Note that we assumed $y=\sqrt{x}$ in the above calculation. Choosing another branch, $y=-\sqrt{x}$, produces the asymptotic expansion of the other solution of the Airy equation, the function ${\rm Bi}(x)$, again up to an overall factor. 
}
\end{example}

\section{Topological recursion for the harmonic oscillator curve}
\label{sect_HO}

In this section we show that an analogue to the Airy case statements holds for the {\it harmonic oscillator curve} $\mathcal L^{H}$
\begin{equation}
\label{HOcurve}
y^2=x^2-c^2,
\end{equation}
where $c\in \mathbb C$ is a constant. As mentioned in the introduction, we quantize the curve by replacing $x,y$ by operators $\hat{x}, \hat{y}$ and choose the quantization prescription $\hat{x}=x$ and $\hat{y}=\hbar \partial/\partial x$ (note that $\hat{y}$ differs from  the conjugate momentum $\hat{p}_x$ of quantum mechanics which must be a Hermitian operator and is given by $-i\hbar \partial/\partial x$). The resulting Schr\"odinger equation is   
\begin{equation}
\label{HOSchrodinger}
\hbar^2\frac{\partial^2}{\partial x^2}\Psi(x,\hbar)=(x^2-c^2)\Psi(x,\hbar).
\end{equation}
We show that the topological recursion of Eynard-Orantin applied to the curve $\mathcal L^{H}$ reproduces the WKB approximation to the solution of the Schr\"odinger equation \eqref{HOSchrodinger}.
Let us introduce local parameters on the curve $\mathcal L^{H}$. To this end, let us bring the curve \eqref{HOcurve} into the form $\tilde{y}^2=\tilde{x}$ by the change of coordinates
\begin{equation*}
\tilde{x}=\frac{x-c}{x+c}\;, \qquad\mbox{and} \qquad \tilde{y}^2=\frac{x-c}{x+c}=\frac{y^2}{(x+c)^2}\;.
\end{equation*}
Then $z=\pm\tilde{y}$ is a local parameter near the point $(\tilde{x}, \tilde{y})=(0,0)$ and $1/z$ is a local parameter near the point at infinity of the curve $\tilde{y}^2=\tilde{x}$. In terms of the initial curve, we choose our local parameter to be
\begin{equation*}
z=\sqrt{\frac{x-c}{x+c}}
\end{equation*}
near the ramification point  at $x=c$ and $1/z$ is the parameter near the second ramification point at $x=-c.$ Let us keep track of the ambiguity of sign using  the factor $\epsilon$ such that $\epsilon^2= 1$. Then the coordinates $x$ and $y$ of the curve are expressed in terms of $z$ as follows. 

\begin{equation}
\label{localparameter}
 x=-c\frac{z^2+1}{z^2-1}, \qquad y = \epsilon\frac{2cz}{z^2-1},\qquad dx=\frac{4czdz}{(z^2-1)^2}.
\end{equation}
Thus we start the topological recursion on the harmonic oscillator curve $\mathcal L^{H}$ \eqref{HOcurve} with the following initial data: 
\begin{equation}
\label{HOinitial}
w^{H}_{0,1}(z) = ydx = \epsilon\frac{8c^2z^2\,dz}{(z^2-1)^3}, \qquad\qquad
w^{H}_{0,2}(z_1,z_2) = \frac{dz_1dz_2}{(z_1-z_2)^2}.
\end{equation}

The involution $*$ acts on the points of $\mathcal L^{H}$ by $(x,y)^*=(x,-y)$ and thus, we  compute the recursion kernel  \eqref{K} for  the harmonic oscillator curve: 
\begin{equation}
\label{HOkernel}
K^{H}(z,z_1)=\frac{\epsilon}{16 c^2}\frac{(1-z^2)^3 \,dz_1}{(z_1^2-z^2)z \,dz}\;.
\end{equation}

The WKB approximation gives a solution to the Schr\"odinger equation \eqref{HOSchrodinger} for the harmonic oscillator in the form of the series \eqref{WKB}. The following theorem states that this series is recovered by the topological recursion on the curve $\mathcal L^{H}$.

\begin{theorem}
\label{thm_WKB}
Let $w_{g,n}^{H}$ be the multi-differentials obtained by the topological recursion \eqref{recursion} on the harmonic oscillator curve $\mathcal L^H$ \eqref{HOcurve}, where $\epsilon=-1$ is chosen in the initial data \eqref{HOinitial}. Let $F_{g,n}^H$ with $2g-2+n>0$ be the functions defined by integrating $w_{g,n}^H$ as follows
\begin{equation}
\label{FgnHO}
F^H_{g,n}(z_1,\dots, z_n) =\frac{1}{2^n} \int^{z_1}_{-z_1} \dots \int^{z_n}_{-z_n} w^H_{g,n}(z_1' \dots z_n').
\end{equation}
 Then the function 
\begin{equation*}
\Psi(x, \hbar) = {\rm exp} \sum_{m=0}^\infty \hbar^{m-1}S_m(x)
\end{equation*}
with
\begin{eqnarray*}
&&S_0(x) = -\frac{1}{2}\int^{(x,y)}_{(x,-y)} y\,dx\;,
\nonumber
\\
&&S_1(x) = -\frac{1}{2}{\rm log}\,y \,,
\nonumber
\\
&&S_m(x) = \sum_{2g+n-2=m-1} \frac{1}{n!} F_{g,n}^{H}(z,\dots,z)  \,, \qquad m\geq 2,
\end{eqnarray*}
satisfies the Schr\"odinger equation \eqref{HOSchrodinger} for the harmonic oscillator. 
 \end{theorem}

{\it Proof.} See Section \ref{sect_proof}.

\begin{remark} The equalities for the  $S_m(x)$  should be understood to hold up to the addition of arbitrary constants since the wave function $\Psi(x,\hbar)$ is only determined by the WKB method up to an overall factor. 
\end{remark}

\begin{remark}
\label{rmk_choice}
Note that the choice $\epsilon=-1$ we make in the theorem corresponds to $y=\sqrt{x^2-c^2}$ according to \eqref{localparameter}. This also agrees with the choice of sign in  $S'_0=-y=-\sqrt{x^2-c^2}$ which we need to make in the WKB method for the harmonic oscillator to obtain a square integrable wave function, see Example \ref{example_Sk}.
\end{remark}

\begin{example}
\label{example_HO}
{\rm Let us give a few examples of the differentials $w^{H}_{g,n}$ and functions $F^H_{g,n}$ defined by \eqref{FgnHO}.  As before, we use the notation $dz_{[k]} = dz_1\dots dz_k$ and  keep both signs in \eqref{HOinitial}.
For $2g-2+n=1$ we have
\begin{eqnarray}
&&w_{0,3}^H(z_1,z_2,z_3)=\epsilon \frac{dz_{[3]}}{2^3c^2}\left(1-\frac{1}{z_1^2z_2^2z_3^2} \right); 
\nonumber
\\
&& w_{1,1}^H(z_1)= \epsilon\frac{(z_1^2-1)^3\,dz_1}{2^6\,c^2\, z_1^4};
\nonumber
\\
&&F^H_{0,3}(z_1,z_2,z_3)= \frac{\epsilon}{2^3c^2}\left(z_1z_2z_3+\frac{1}{z_1z_2z_3} \right);
\nonumber
\\
&&F^H_{1,1}(z_1)=\frac{\epsilon}{2^6\,c^2} \left( \frac{z_1^3}{3} -3z_1 -\frac{3}{z_1} + \frac{1}{3z_1^3}\right).
\label{wgn12}
\end{eqnarray}
For $2g-2+n=2$ we have
\begin{equation*}
 w_{0,4}^H (z_1,z_2,z_3,z_4)= \frac{dz_{[4]}}{2^6c^4}
\left(
\frac{3}{ z_1^2z_2^2z_3^2z_4^2} \sum_{i=1}^4\frac{1}{z_i^2}  - \frac{9}{z_1^2z_2^2z_3^2z_4^2}  -\sum_{i<j}\frac{1}{z_i^2 z_j^2} - 9 + 3\sum_{i=1}^4 z_i^2
\right);
\end{equation*}
\begin{multline*}
w_{1,2}^H(z_1,z_2) = \frac{dz_{[2]}}{2^9 c^4} \left( \frac{5}{ z_1^2z_2^6} +\frac{3}{ z_1^4 z_2^4} + \frac{5}{ z_1^6z_2^2}
-\frac{18}{z_1^2z_2^4}-\frac{18}{z_1^4z_2^2} +\frac{27}{z_1^2z_2^2}-\frac{4}{z_1^2}-\frac{4}{z_2^2} \right.
\\
\left. + 27
-18z_1^2-18z_2^2 +5z_1^4+3z_1^2z_2^2+5z_2^4
 \right).
\end{multline*}
}
\end{example}

\vskip 0.5cm

\begin{remark} We note that some terms in the expressions for the multi-differentials $w_{g,n}^H$ in this example combine to produce the corresponding differentials computed for the Airy curve. In general, we observe that the $w_{g,n}^H$ in the stable range $2g-2+n>0$ have the form:

\begin{multline*}
w_{g,n}^H(p_1,\dots, p_n) =  \left(- \frac{ \epsilon}{8c^2} \right)^{2g-2+n} \!\!\!\!\!\!\!\!\!\!\!\!\!\!\!\!\!\!\sum_{\substack{a_1,\dots, a_n\\a_1+\dots+a_n=3g-3+n, \; a_i\geq 0}}
\!\!\!\!\!\!\!\!\!\!\!\!\!
  \langle \tau_{a_1}\dots\tau_{a_n} \rangle_g       \left( \prod_{i=1}^n\frac{(2a_i+1)!!}{p_i^{2a_i+2}} + (-1)^n\prod_{i=1}^n{(2a_i+1)!!}{p_i^{2a_i}}  \right) 
\\
+\sum_{k\geq1} \sum_{\substack{a_1,\dots, a_n\\a_1+\dots+a_n=3g-3+n-k}}\frac{ C^g_{a_1\dots a_n}}{c^{4g-4+2n}
} \left( \prod_{i=1}^n\frac{1}{p_i^{2a_i+2}} + (-1)^n\prod_{i=1}^n{p_i^{2a_i}}  \right) 
\end{multline*}
with some rational coefficients $C^g_{a_1\dots a_n}$.

\end{remark}

 %\begin{remark}
 In Section \ref{sect_WKB} we showed that the energies of the harmonic oscillator  are obtained from the energy quantization condition \eqref{quant} where the integration contour goes in the $x$-plane around the two turning points $x=\pm c$. 
This integral can be interpreted as the integral of the differential $\hbar^{m-1} S_m'(x) \, dx$ over the contour $\gamma$ on the harmonic oscillator curve encircling two ramification points $(-c,0)$ and $(c,0)$  in the counterclockwise direction. This contour lies on the sheet of the double covering of the $x$-sphere specified by the condition $y=\sqrt{x^2-c^2}$, which corresponds to  the choice we make in  the WKB method, see Remark \ref{rmk_choice}.

 Given that all $S_m$ of the harmonic oscillator for $m\geq 2$ are single valued functions on the curve (see for instance  Example \ref{example_Sk}), the integral of $S_m'(x) \, dx$ over a closed contour is zero for $m\geq 2$. Therefore the quantization condition \eqref{quant} reduces to
 \begin{equation*}
\frac{1}{2 \pi \i} \oint_\gamma  \left( \frac{S_0'(x)}{\hbar} + S'_1(x)\right) \, dx  = n.
\end{equation*} 
From Example \ref{example_Sk} we get that the differential under the integral is given by 
 \begin{equation*}
\left( \frac{S_0'(x)}{\hbar} + S'_1(x)\right) \, dx  
%= \frac{y\,dx}{\hbar} - \frac{1}{2}\frac{x\,dx}{y^2} 
= \left(-\frac{y}{\hbar} - \frac{1}{2}\frac{x}{y^2}\right)\,dx.
\end{equation*} 

 As is easy to see the integral over the contour $\gamma$ is equal to the $-2\pi\i$ times the residue at the point at infinity on the sheet where $y=\sqrt{x^2-c^2}$, let us denote this point $\infty^+$. Thus the above energy quantization condition becomes 
 \begin{equation*}
-\,\underset{\infty^{+}}{\rm Res} \,\left(-\frac{y}{\hbar} - \frac{1}{2}\frac{x}{y^2}\right)\,dx  =n.
\end{equation*} 
Calculating the residue in the local parameter $\tau=1/x$ near $\infty^+$ we have
 \begin{equation*}
\underset{\infty^{+}}{\rm Res} \,y\,dx  = \underset{\infty^{+}}{\rm Res} \,(\sqrt{x^2-c^2})\,dx =\frac{c^2}{2} \qquad\mbox{and}\qquad
\underset{\infty^{+}}{\rm Res} \,\frac{x\,dx}{y^2} = -1.
\end{equation*} 
Thus the energy quantization condition yields
 \begin{equation*}
\frac{c^2}{2}  = \left(n + \frac{1}{2}\right)\hbar \, ,
\end{equation*} 
which, given that $c^2=\frac{2}{\omega}E$, reproduces the possible energy levels $E_n$ according to \eqref{eho}. Note that our calculation implies that $$\frac{1}{2 \pi \i} \oint_\gamma  S'_1(x) \, dx=\frac{1}{2}\underset{\infty^{+}}{\rm Res} \,\frac{x\,dx}{y^2}=-\frac{1}{2}\,,$$ which confirms our general statement about the contribution of $S'_1\,dx$ made in the discussion after formula \eqref{quant}.

% \end{remark}

\section{Proof of Theorem \ref{thm_WKB}}
\label{sect_proof}

To prove that the topological recursion procedure reproduces the WKB approximation, we use the ideas of Zhou \cite{Zhou2}. First, let us state the following technical lemma the proof for which is given in the appendix. 

\begin{lemma}
\label{lemma_residues}
For the recursion kernel $K^H$ \eqref{HOkernel} of the harmonic oscillator curve, we have
\begin{equation}
\hspace{-7.5 cm}
\left(\underset{z=0}{\rm Res}+\underset{z=\infty}{\rm Res} \right) K^{H}(z,z_1) \frac{(dz)^2}{z^{2k}} =\frac{\epsilon}{16 c^2}\frac{(1-z_1^2)^3}{z_1^{2k+2}} dz_1.
\label{res1}
\end{equation}

\begin{multline}
\left(\underset{z=0}{\rm Res}+\underset{z=\infty}{\rm Res} \right) K^{H}(z,z_1)\left( - \frac{2z^2+2z_j^2}{(z^2-z_j^2)^2}  \right) \frac{(dz)^2}{z^{2k}}= 
\\
=- \frac{\epsilon \, dz_1}{8c^2} \left\{ \frac{(1-z_1^2)^3}{z_1^6} \sum_{s=0}^{1-k} (2s+1)z_j^{2s}z_1^{2(1-k-s)} + \frac{(1-z_1^2)^3}{z_j^2z_1^2} \sum_{t=0}^{k-3} \frac{2t+1}{z_j^{2t} z_1^{2(k-t)}} \right.
\\
\left. + \frac{1}{z_j^2z_1^2}\left(  \frac{(2k-3)(1-3z_1^2 + 3z_1^4)}{z_j^{2(k-2)} z_1^4} + \frac{(2k-1)(1-3z_1^2)}{z_j^{2(k-1)}z_1^2} + \frac{2k+1}{z_j^{2k}}  \right) \right\}
.
\label{res2}
\end{multline}

\end{lemma}

The reason we need the residues from Lemma \ref{lemma_residues} is the following. 
Note that the recursion  formula \eqref{recursion} forbids the unstable indices $(g,n)=(0,1)$ in the sum over all splittings $g_1+g_2=g$ but allows the unstable indices $(g,n)=(0,2)$. Let us rewrite the formula while separating the ``unstable'' terms from the rest of the sum. Namely, we have
\begin{multline*}
w^H_{g,n}(z_1,\dots, z_n) =\left(\underset{z=0}{\rm Res}+\underset{z=\infty}{\rm Res} \right) K^{H}(z,z_1){\Big\{} w^H_{g-1,n+1}(z,-z, z_2,\dots, z_n) 
%\right.
\\
%\left.
+ \sum_{\substack{g_1+g_2=g\\I\coprod J=\{2,\dots, n\}}}^{\mbox{stable }} w^H_{g_1, |I|+1}(z,z_I)w^H_{g_2, |J|+1}(-z,z_J)
+\sum_{k=2}^n w^H_{g,n-1}(z, z_{[n]_{1,k}}) \left(w_{0,2}(-z,z_k) - w_{0,2}(z,z_k)\right)
{\Big\}},
%\right\}.
\end{multline*}
where the notation $[n]_{1,k}$ for the set of numbers from $2$ to $n$ skipping the number $k$ is used. We also used the relation $w^H_{g,n-1}(-z, z_{[n]_{1,k}})=-w^H_{g,n-1}(z, z_{[n]_{1,k}})$ which we explain in Corollary \ref{corollary_wgn}, after rewriting the above recursion formula by computing the difference $(w_{0,2}(-z,z_k) - w_{0,2}(z,z_k))$ in terms of the coordinate $z$ using \eqref{HOinitial}: 

\begin{multline}
\label{HOrecursion}
w^H_{g,n}(z_1,\dots, z_n) = \left(\underset{z=0}{\rm Res}+\underset{z=\infty}{\rm Res} \right) K^{H}(z,z_1){\Big\{} w^H_{g-1,n+1}(z,-z, z_2,\dots, z_n)
%\right.
\\
%\left.
 + \sum_{\substack{g_1+g_2=g\\I\coprod J=\{2,\dots, n\}}}^{\mbox{stable } } w^H_{g_1, |I|+1}(z,z_I)w^H_{g_2, |J|+1}(-z,z_J)-\sum_{k=2}^n w^H_{g,n-1}(z, z_{[n]_{1,k}}) \frac{2(z^2+z_k^2)}{(z^2-z_k^2)^2}   dz \, dz_k
{\Big\}}  \,.
\end{multline}

\begin{corollary}
\label{corollary_wgn}
The differentials $w_{g,n}^H(z_1,\dots, z_n)$ for the indices of the stable type $2g-2+n>0$ are Laurent polynomials with only even exponents of $z_i$.
\end{corollary}

{\it Proof.}
From Example \ref{example_HO} we know that the statement of the corollary holds for $w_{1,1}^H$ and $w_{0,3}^H$, that is the first generation of the differentials in the stable range, those for which $2g-2+n=1$. 
For the differentials $w_{g,n}$ with $2g-2+n>1$, the recursion formula \eqref{HOrecursion} does not contain the ``non-stable'' differentials $w_{0,1}$ and $w_{0,2}$ and therefore the ``stable'' $w_{g,n}$ are obtained recursively by computing the residues from Lemma \ref{lemma_residues} of Laurent polynomials with only even exponents of $z_i$.
$\Box$

The  functions $F^H_{g,n}(z_1,\dots, z_n)$ with $2g-2+n>0$ defined in Theorem \ref{thm_WKB} due to Corollary \ref{corollary_wgn} and the fact, see \cite{EO}, that the $w^H_{g,n}$ are symmetric in all variables  have the general form:
\begin{equation*}
F^H_{g,n}(z_1,\dots, z_n)=  \sum_{a_1,\dots, a_n} c_{a_1,\dots,a_n}^g\prod_{i=1}^n \frac{1}{z_i^{2a_i+1}}
\end{equation*}
where the exponents $a_i$ can be both positive and negative.
As a corollary of the recursion \eqref{HOrecursion}, the antiderivatives $F^H_{g,n}$ also satisfy a recursion formula obtained in the next lemma. This recursion is a particular case of the ``differential recursion'' for the free energies derived in \cite{DumitrescuMulase}. This is a key lemma in our proof of Theorem \ref{thm_WKB}.

\begin{lemma} The functions $F^H_{g,n}$ with $2g-2+n>1$ obtained by the integration of the $w_{g,n}^H$ as in \eqref{FgnHO} satisfy the following  recursion 
\label{lemma_diff_recursion}
\begin{multline*}
\partial_{s} F^H_{g,n}(s,z,\dots, z)\Big{|}_{s=z} = -\frac{\epsilon}{16 c^2}\frac{(1-z^2)^3}{z^2}  \partial_s\partial_t F^H_{g-1,n+1}(s,t,z, \dots, z)\Big{|}_{s=t=z} 
\\
- \frac{\epsilon}{16 c^2}\frac{(1-z^2)^3}{z^2}  \sum_{\substack{g_1+g_2=g\\n_1+n_2=n-1}}^{\mbox{\rm stable } } \frac{(n-1)!}{n_1!n_2!}  \partial_{s}F^H_{g_1, n_1+1}(s,z, \dots, z)\partial_{s}F^H_{g_2, n_2+1}(s,z, \dots, z)\Big{|}_{s=z}
\\
 + \frac{\epsilon}{16 c^2} \frac{z^2}{(z^2-1)^3} (n-1) \left(  \frac{(s^2-1)^3}{s^2} \frac{d}{ds} \right)^2 F^H_{g,n-1}(s, z, \dots, z)\Big{|}_{s=z} .
\end{multline*}
The initial condition for this recursion are the functions $F^H_{1,1}$ and $F^H_{0,3}$ given in \eqref{wgn12}. 
\end{lemma}

{\it Proof.} The first step is to rewrite the recursion formula  for $w^H_{g,n}$ in a polynomial form, that is to compute the residues in \eqref{HOrecursion}. To this end and following \cite{Zhou2},
let us introduce an operator $D_{z_1,z_j}$ acting on even powers of $s$ (and not acting on any $z_i$) as follows ($k\in\mathbb Z$): 
\begin{multline}
D_{z_1,z_j}\frac{ds}{s^{2k}} 
= \left\{ \frac{(1-z_1^2)^3}{z_1^6} \sum_{r=0}^{1-k} (2r+1)z_j^{2r}z_1^{2(1-k-r)} + \frac{(1-z_1^2)^3}{z_j^2z_1^2} \sum_{t=0}^{k-3} \frac{2t+1}{z_j^{2t} z_1^{2(k-t)}} \right.
\\
\left. + \frac{1}{z_j^2z_1^2}\left(  \frac{(2k-3)(1-3z_1^2 + 3z_1^4)}{z_j^{2(k-2)} z_1^4} + \frac{(2k-1)(1-3z_1^2)}{z_j^{2(k-1)}z_1^2} + \frac{2k+1}{z_j^{2k}}  \right) \right\}dz_1dz_j. \label{operd}
\end{multline}
As is easy to see, this is proportional to  the second residue from Lemma \ref{lemma_residues}. 
Using this operator and the formulas from Lemma \ref{lemma_residues}, due to Corollary \ref{corollary_wgn} the recursion \eqref{HOrecursion} for $w^H_{g,n}$ rewrites as follows: 
\begin{multline}
\label{wgnrecur}
w^H_{g,n}(z_1,\dots, z_n) = -\frac{\epsilon}{16 c^2}\frac{(1-z_1^2)^3}{z_1^2} w^H_{g-1,n+1}(z_1,z_1, z_2,\dots, z_n)\frac{1}{dz_1} 
\\
-\frac{\epsilon}{16 c^2}\frac{(1-z_1^2)^3}{z_1^2}\sum_{\substack{g_1+g_2=g\\I\coprod J=\{2,\dots, n\}}}^{\mbox{stable } } w^H_{g_1, |I|+1}(z_1,z_I)w^H_{g_2, |J|+1}(z_1,z_J)\frac{1}{dz_1}
%\\
-\frac{\epsilon}{8c^2}\sum_{j=2}^n D_{z_1,z_j} w^H_{g,n-1}(s, z_{[n]_{1,j}}) . 
\end{multline}
We now want to integrate this recursion in order to produce a recursion formula for $F^H_{g,n}$. Note that $F^H_{g,n}$ are only defined as integrals of $w^H_{g,n}$ for the indices in the stable range, that is $2g-2+n>0$. To avoid appearances of $w^H_{0,1}$ and $w^H_{0,2}$ in \eqref{wgnrecur} we see that we need to restrict the indices to those satisfying $2g-2+n>1$. By integrating \eqref{wgnrecur} with respect to $z_2,\dots, z_n$ as in \eqref{FgnHO}, that is from $-z_i$ to $z_i$ for every $i\geq2$, and then dividing by $2^{n-1}$ and by $dz_1$ we get the recursion  in terms of the $F^H_{g,n}$, now for the indices $(g,n)$ satisfying $2g-2+n>1$:
\begin{multline}
\label{HOFrecursion}
\partial_{z_1} F^H_{g,n}(z_1,z_2,\dots, z_n) = -\frac{\epsilon}{16 c^2}\frac{(1-z_1^2)^3}{z_1^2}  \partial_s\partial_t F^H_{g-1,n+1}(s,t,z_2, \dots, z_n)\Big{|}_{s=t=z_1} 
\\
- \frac{\epsilon}{16 c^2}\frac{(1-z_1^2)^3}{z_1^2}  \sum_{\substack{g_1+g_2=g\\I\coprod J=\{2,\dots, n\}}}^{\mbox{stable } } \partial_{z_1}F^H_{g_1, |I|+1}(z_1,z_I)\partial_{z_1}F^H_{g_2, |J|+1}(z_1,z_J)
%\\
- \frac{\epsilon}{8c^2}\sum_{j=2}^n \mathcal D_{z_1,z_j} \partial_s F^H_{g,n-1}(s, z_{[n]_{1,j}})\, ,
\end{multline}
where $ \mathcal D_{z_1,z_j}$ is a new operator acting on even powers of $s$ (but not acting on any $z_i$) defined by
\begin{multline}
\mathcal D_{z_1,z_j}  \frac{1}{s^{2k}} := \frac{1}{2} \int^{z_j}_{-z_j} D_{z_1,z_j'} \frac{ds}{s^{2k} } \frac{1}{dz_1} \\
= 
\left\{ \frac{(1-z_1^2)^3}{z_1^6} \sum_{r=0}^{1-k}z_j^{rs+1}z_1^{2(1-k-r)} - \frac{(1-z_1^2)^3}{z_j z_1^2} \sum_{t=0}^{k-3} \frac{1}{z_j^{2t} z_1^{2(k-t)}} \right.
\\
\left. - \frac{1}{z_j z_1^2}\left(  \frac{1-3z_1^2 + 3z_1^4}{z_j^{2(k-2)} z_1^4} + \frac{1-3z_1^2}{z_j^{2(k-1)}z_1^2} + \frac{1}{z_j^{2k}}  \right) \right\} .
\label{curlyd}
\end{multline}

Now we need to set all variables equal $z_i=z$ in \eqref{HOFrecursion}.  To do so, let us first look at what happens with $\mathcal D_{z_1,z_j} \partial_s \frac{1}{s^{2a-1}}$. A quantity of this form is present in every term of the last sum in \eqref{HOFrecursion}. A straightforward calculation yields
\begin{equation}
\label{temp}
\mathcal D_{z_1,z_j} \partial_s \frac{1}{s^{2a-1}}\Big{|}_{z_1=z_j=z} =\mathcal D_{z_1,z_j} \frac{-(2a-1)}{s^{2a}}\Big{|}_{z_1=z_j=z}
   =-\frac{(2a-1)(z^2-1)^2}{z^{2a+3}} \left\{ (a-2)z^2-a-1  \right\} .
\end{equation}

On the other hand, we can obtain the expression in the right hand side of \eqref{temp} in another way. Consider
\begin{equation*}
\left(  \frac{(z^2-1)^3}{z^2} \frac{d}{dz} \right)^2 \frac{1}{z^{2a-1}} =2(2a-1)\frac{(z^2-1)^5}{z^{2a+5}} \left\{ (a-2)z^2-a-1 \right\}.
\end{equation*}
Therefore
\begin{equation*}
\mathcal D_{z_1,z_j} \partial_s \frac{1}{s^{2a-1}}\Big{|}_{z_1=z_j=z} = -\frac{z^2}{2(z^2-1)^3}\left(  \frac{(z^2-1)^3}{z^2} \frac{d}{dz} \right)^2 \frac{1}{z^{2a-1}} .
\end{equation*}

Using this relation and setting $z_i=z$ for all $i$ in \eqref{HOFrecursion}, we prove the lemma. $\Box$

\vskip 1cm

Recall that we want to show that the coefficients $S_m$ for $m\geq 2$ in the WKB approximation to the wave function $\Psi$   can be obtained from the functions $F^H_{g,n}$ as stated in  Theorem   \ref{thm_WKB}. Let us therefore look at the WKB approximation for the harmonic oscillator \eqref{HOSchrodinger}. 
In this case we have, see Remark \ref{rmk_choice},
\begin{eqnarray*}
\label{S0prime}
&&S'_0=-y=-\sqrt{x^2-c^2},
\\
\nonumber
&& S'_1=-\frac{1}{2}({\rm \log}\,y)'=-\frac{1}{2}({\rm \log}\sqrt{x^2-c^2})'= -\frac{1}{2}\frac{x}{x^2-c^2} \,.
\end{eqnarray*}
Upon integration, we recover the $S_0(x)$ and $S_1(x)$ of  Theorem   \ref{thm_WKB} up to irrelevant additive constants.
Consider now $S_2$ which is given in  Example  \ref{example_Sk}:
\begin{equation}
\label{S2z}
S_2
= \frac{ x^3-6c^2x }{24c^2y^3} 
=\frac{5-9z^2 -9z^4+5z^6}{192\,\epsilon\,c^2\,z^3}  \, .
\end{equation}
Using  results from Example \ref{example_HO}, one easily verifies that, as stated in Theorem   \ref{thm_WKB}   (recall that according to the theorem we choose $\epsilon=-1$), 
\begin{equation}
\label{S2F}
S_2(z)=F^H_{1,1}(z) +\frac{1}{3!} F^H_{0,3}(z,z, z). 
\end{equation}

We now turn our attention to the differential recursion equation \eqref{WKB_WKB} for  the $S_m'$  for $m \geq 3$ which we rewrite, plugging in the values in terms of $x$ and $y$ for $S'_0$ and $S'_1$, in the form
\begin{equation}
\label{WKBHO_1}
S_m'=\frac{1}{ 2y }\left(S''_{m-1}-\frac{x}{x^2-c^2}S_{m-1}'+\sum_{\substack{i+j=m\\i,j\geq 2}} S'_i S'_j\right).
\end{equation}
Here the prime denotes a derivative with respect to $x$.
Now we want to rewrite \eqref{WKBHO_1} using differentiation with respect to $z$. According to the relationship \eqref{localparameter} between $x,y$ and $z$ we have 
\begin{equation*}
\frac{d}{dx}=\frac{dz}{dx}\frac{d}{dz}=\frac{(z^2-1)^2}{4cz}\frac{d}{dz}
\qquad\mbox{and}\qquad 
y=\epsilon\frac{2cz}{z^2-1}.
\end{equation*}
As explained in Remark \ref{rmk_choice}, we must choose $\epsilon=-1$ and thus \eqref{WKBHO_1} becomes
\begin{multline*}
\frac{(z^2-1)^2}{4cz}\frac{d}{dz}S_m=
-\frac{z^2-1}{ 4cz } \Bigg( \frac{(z^2-1)^3(2z^2+1)}{8c^2z^3}\frac{d}{dz}S_{m-1}+
\\
 + \frac{(z^2-1)^4}{16c^2z^2}\frac{d^2}{dz^2}S_{m-1}
+\frac{(z^2-1)^4}{16c^2z^2}\sum_{\substack{i+j=m\\i,j\geq 2}} \frac{d}{dz}S_i \frac{d}{dz}S_j\Bigg). 
\end{multline*}
Note that the first two terms on the right can be rewritten as follows
\begin{equation*}
\frac{(z^2-1)^3(2z^2+1)}{8c^2z^3}\frac{d}{dz} + \frac{(z^2-1)^4}{16c^2z^2}\frac{d^2}{dz^2}=\frac{z^2}{16c^2(z^2-1)^2}\left(  \frac{(z^2-1)^3}{z^2} \frac{d}{dz} \right)^2
\end{equation*}
and therefore the differential recursion relation we obtain from  WKB for $S_m$, $m\geq 3$, is
\begin{equation}
\label{WKB_3}
\frac{d}{dz}S_m= -\frac{z^2}{16c^2(z^2-1)^3}\left(  \frac{(z^2-1)^3}{z^2} \frac{d}{dz} \right)^2S_{m-1} -\frac{(z^2-1)^3}{16c^2z^2}\sum_{\substack{i+j=m\\i,j\geq 2}} \frac{d}{dz}S_i \frac{d}{dz}S_j 
\end{equation}
with initial condition $S_2(z)$ given by \eqref{S2z}.

Recall that the  claim of Theorem \ref{thm_WKB} is that $S_m(z)=\sum_{2g+n-1=m} \frac{1}{n!} F^H_{g,n}(z,\dots, z) $ for $m\geq 2$. In the next lemma we suppose this true for $m\geq 3$ and see how the WKB differential equation \eqref{WKB_3} may be rewritten  in terms of $F^H_{g,n}.$ Note that the $S_m(z)$, $m\geq 2,$ are Laurent polynomials because $S_2$ is, see \eqref{S2z}, and because of the form of the recursion \eqref{WKB_3}. Therefore each $S_m$, $m\geq 2,$ can be written in the form assumed in the next lemma.

\begin{lemma}
\label{lemma_comp}
Let $V_{g,n}(z_1,\dots,z_n)$ be symmetric Laurent polynomials of $n$ arguments  such that 
\begin{equation*}
S_m(z) =\sum_{2g+n-2=m-1} \frac{1}{n!} V_{g,n}(z,\dots, z) \quad\mbox{for} \quad m\geq 2. 
\end{equation*}
Then the differential equation \eqref{WKB_3} is equivalent to the following equation in terms of $V_{g,n}$
\begin{multline}
\label{WKB_Vgn}
\sum_{2g+n-2=m-1} \frac{1}{(n-1)!} \partial_s V_{g,n}(s,z,\dots, z)\Big{|}_{s=z} 
\\
=-\frac{(z^2-1)^3}{16c^2z^2} \sum_{2 g+ n-2=m-1} \frac{1}{( n-1)!}  \partial_s \partial_tV_{ g-1, n+1}(s,t,z,\dots, z)\Big{|}_{s=t=z} 
\\
-\frac{z^2}{16c^2(z^2-1)^3} \sum_{2 g+ n-2=m-1} \frac{1}{( n-2)!} \left(\frac{(s^2-1)^3}{s^2} \partial_s \right)^2V_{ g, n-1}(s,z,\dots, z)\Big{|}_{s=z}
\\
-\frac{(z^2-1)^3}{16c^2z^2}  
\sum_{2g+n-2=m-1} \sum_{\substack{g_1+g_2=g\\n_1+n_2=n+1}}^{\mbox{\rm stable}}
 \frac{1}{(n_1-1)!} \partial_s V_{g_1,n_1}(s,z,\dots, z) \frac{1}{(n_2-1)!} \partial_s V_{g_2,n_2}(s,z,\dots, z) \Big{|}_{s=z}.
\end{multline}

\end{lemma}

{\it Proof.} The proof is a straightforward computation. Using the symmetry of the functions $V_{g,n}$, we get directly from \eqref{WKB_3}

\begin{multline*}
\sum_{2g+n-2=m-1} \frac{1}{(n-1)!} \partial_s V_{g,n}(s,z,\dots, z)\Big{|}_{s=z} 
\\
=-\frac{z^2}{16c^2(z^2-1)^3} \sum_{2g+n-2=m-2} \frac{n(n-1)}{n!} \frac{(s^2-1)^3}{s^2} \partial_s \frac{(t^2-1)^3}{t^2} \partial_tV_{g,n}(s,t,z,\dots, z)\Big{|}_{s=t=z} 
\\
-\frac{z^2}{16c^2(z^2-1)^3} \sum_{2g+n-2=m-2} \frac{n}{n!} \left(\frac{(s^2-1)^3}{s^2} \partial_s\right)^2V_{g,n}(s,z,\dots, z)\Big{|}_{s=z}
\\
-\frac{(z^2-1)^3}{16c^2z^2} \sum_{\substack{i+j=m\\i,j\geq 2}} \left(\sum_{2g_1+n_1-2=i-1} \frac{n_1}{n_1!} \partial_sV_{g_1,n_1}(s,z,\dots, z)\right)\left(\sum_{2g_2+n_2-2=j-1} \frac{n_2}{n_2!} \partial_sV_{g_2,n_2}(s,z,\dots, z)\right) \Big{|}_{s=z}.
\end{multline*}

Resumming the terms in the  last line we arrive at
\begin{multline*}
\sum_{2g+n-2=m-1} \frac{1}{(n-1)!} \partial_s V_{g,n}(s,z,\dots, z)\Big{|}_{s=z} 
\\
=-\frac{(z^2-1)^3}{16c^2z^2} \sum_{2g+n-2=m-2} \frac{1}{(n-2)!}  \partial_s \partial_sV_{g,n}(s,t,z,\dots, z)\Big{|}_{s=t=z} 
\\
-\frac{z^2}{16c^2(z^2-1)^3} \sum_{2g+n-2=m-2} \frac{1}{(n-1)!} \left(\frac{(s^2-1)^3}{s^2} \partial_s\right)^2V_{g,n}(s,z,\dots, z)\Big{|}_{s=z}
\\
-\frac{(z^2-1)^3}{16c^2z^2}  
 \sum_{2g+n-2=m-1} \sum_{\substack{g_1+g_2=g\\n_1+n_2=n+1}}^{\mbox{stable}}
 \frac{1}{(n_1-1)!} \partial_sV_{g_1,n_1}(s,z,\dots, z) \frac{1}{(n_2-1)!} \partial_sV_{g_2,n_2}(s,z,\dots, z) \Big{|}_{s=z}.
\end{multline*}

This coincides with the claim of the lemma up to the following change of the summation variables $g$ and $n$ in some of the sums. In the first sum in the right hand side we put $g=\tilde g-1$ and $n=\tilde n+1. $ In the second sum we put $\tilde g=g$ and $ n=\tilde n-1$.
$\Box$

To finish the proof of Theorem \ref{thm_WKB} we note that if we divide both sides of the  recursion formula for $F^H_{g,n}$ obtained in Lemma \ref{lemma_diff_recursion} by ${(n-1)!}$, set $\epsilon=-1$  and sum over pairs $(g,n)$ such that $2g+n-2=m-1$, we obtain   \eqref{WKB_Vgn} from Lemma \ref{lemma_comp} if we put $V_{g,n}=F^H_{g,n}$. This shows that the quantities 
\begin{equation}
\label{claim}
S_m(z)=\sum_{2g+n-2=m-1} \frac{1}{n!} F^H_{g,n}(z,\dots, z)  + const
\end{equation}
satisfy the differential recursion \eqref{WKB_3} of  WKB for $m\geq 3.$ Since we have also checked that \eqref{claim} is satisfied by the initial data of the two recursions, that is $S_2(z)=F^H_{1,1}(z) +\frac{1}{3!} F^H_{0,3}(z,z, z)$, see \eqref{S2z} and \eqref{S2F},
this completes our proof of Theorem \ref{thm_WKB}. $\Box$

\begin{remark}
Note that the WKB method determines $S_m$ only up to an additive constant, see for example equation \eqref{WKB_3}. In other words, we proved that equation \eqref{claim} is satisfied up to addition of an arbitrary constant. 
 It turns out that for even indices $m$ equation \eqref{claim} is exactly satisfied,  and for odd $m$ it is satisfied up to a constant. We can see from Examples  \ref{example_Sk} and \ref{example_HO}  that 
$$S_3(z) =  \frac{1}{4!}F_{0,4} + \frac{1}{2!}F_{1,2}  -\frac{1}{32 \,c^4}.$$ 
\end{remark}

\section{Relationship with the space of metric ribbon graphs}
\label{sect_ribbon}

In this section we show the relationship between the harmonic oscillator curve and the results of \cite{Chapman, MulasePenkava} concerning the orbifolds $RG_{g,n}$ of metric ribbon graphs of topological type $(g,n)$ with vertices of degree at least three: 

\begin{equation*}
RG_{g,n} = \underset{\substack{\Gamma\; \rm{ribbon \; graph}\\\rm{ \;of \;type} \;(g,n)}}\coprod\frac{\mathbb R_+^{e(\Gamma)}}{{\rm Aut}(\Gamma)} .
\end{equation*}
Here $e(\Gamma)$ is the number of edges of the ribbon graph $\Gamma$, and ${\rm Aut}(\Gamma)$ is the group of ribbon graph automorphisms of $\Gamma$ preserving every face. 
This orbifold is isomorphic to $\mathcal M_{g,n}\times \mathbb R^n_+$. 

It turns out that the differentials $w^H_{g,n}$ for $2g-2+n>0$ contain information about the so-called Poincar\'e polynomials associated with the orbifolds  $RG_{g,n}$. 
The Poincar\'e polynomial of  $RG_{g,n}$ as defined in \cite{MulasePenkava} is given by
\begin{equation}
\label{Poincare1}
P_{g,n}(z_1,\dots, z_n) =   \underset{\substack{\Gamma\; \rm{ribbon \; graph}\\\rm{ \;of \;type} \;(g,n)}}\sum \frac{(-1)^{e(\Gamma)}}{|{\rm Aut}\Gamma|} 
 \underset{\substack{\eta\; \rm{edge}\\\rm{ \;of } \;\Gamma}}\prod \frac{(z_{i_\eta}+1)(z_{j_\eta}+1)}{2(z_{i_\eta}+z_{j_\eta})},
\end{equation}
where ${i_\eta}$, ${j_\eta}$ are the two faces (possibly the same) incident to the edge $\eta$, and $z_{i_\eta}$ $z_{j_\eta}$ are variables assigned to the faces.

It turns out that  
\begin{equation}
\label{harmonicPoincare}
\left(-\frac{c^2}{2\epsilon}\right)^{{2g-2+n}} w^H_{g,n}(z_1,\dots, z_n) =d_{z_1}\dots d_{z_n}P_{g,n}(z_1,\dots, z_n)
\end{equation}
for the pairs $(g,n)$ subject to the stability condition $2g-2+n>0$. This was shown in \cite{Chapman} in the following way. Let us set $c^2=-2\epsilon$ to make the overall factor in \eqref{harmonicPoincare} disappear. Our multi-differentials $w^H_{g,n}$ with this choice of $c^2$ were obtained  in \cite{Chapman} by using the Eynard-Orantin type topological recursion on a different curve defined by 
\begin{equation}
\label{othercurve}
\tilde{x}\tilde{y}=\tilde{y}^2+1.
\end{equation}
By the change of variables
\begin{equation*}
{x}=c\frac{\tilde{x}}{2},\qquad \qquad {y}=c\left(\tilde{y}-\frac{\tilde{x}}{2}\right)
\end{equation*}
this curve is transformed to the harmonic oscillator curve ${y}^2={x}^2-c^2$, so the two curves are equivalent as algebraic curves. The parameter $t$ used in \cite{Chapman} corresponds exactly to our parameter $z$. However, in the topological recursion of Eynard-Orantin, we define the initial differential for ``unstable" indices $(0,1)$ as $w_{0,1}=ydx$ which is not the same for the two curves, $ydx\neq\tilde{y}d\tilde{x}$. The second initial differential $w_{0,2}$  should however be the same, see \eqref{HOinitial}, for both curves if  the original definition of Eynard-Orantin \cite{EO}  is used.
Therefore, the topological recursion procedure of \cite{EO} described in Section \ref{sect_recursion} does not give the same differentials $w_{g,n}$ for the two curves. 

The reason why the multi-differentials $w_{g,n}$ obtained in \cite{Chapman} for the curve
\eqref{othercurve} coincide with our $w^H_{g,n}$ is the following. The point of departure in \cite{Chapman} are the functions $ L_{g,n}(\omega_1,\dots, \omega_n)$ obtained as Laplace transform of the numbers $N_{g,n}(p_1,\dots, p_n)$ representing the number of ribbon graphs of the topological type $(g,n)$ with integer perimeter lengths $(p_1,\dots, p_n)\in\mathbb Z_+^n$ assigned to their faces:
\begin{equation*}
L_{g,n}(\omega_1,\dots, \omega_n) = \sum_{(p_1,\dots, p_n)\in\mathbb Z_+^n} N_{g,n}(p_1,\dots, p_n)\,{\rm e}^{-p_1\omega_1-\dots - p_n\omega_n} \, .
\end{equation*}

By putting the $L_{g,n}$ on the curve using the following correspondence of variables
\begin{equation}
\label{change}
{\rm e}^{-\omega}=\frac{z+1}{z-1}
\qquad\mbox{and}\qquad
{\rm e}^{-\omega_j}=\frac{z_j+1}{z_j-1}
\end{equation}
and differentiating the functions $ L_{g,n}$ with respect to the $z_j$ one obtains the multi-differentials $\mathcal L_{g,n}$:
\begin{equation*}
\mathcal L_{g,n}(z_1,\dots, z_n)=d_{z_1}\dots d_{z_n} L_{g,n}(\omega_1,\dots, \omega_n).
\end{equation*}
It is these differentials that coincide with ours: $\mathcal L_{g,n}=w^H_{g,n}$ (if we keep the constant $c$ arbitrary, then they coincide up to a factor: $\mathcal L_{g,n}=\left({-c^2}/{2\epsilon}\right)^{{2g-2+n}}w^H_{g,n}$). The reason in \cite{Chapman} to consider the topological recursion on the curve \eqref{othercurve} was in order to obtain of $\mathcal L_{g,n}$. Therefore, the differential $w_{0,2}$ was chosen to match $\mathcal L_{0,2}$ which was computed independently of the topological recursion procedure by computing the number of integral ribbon graphs of genus zero having two faces of equal integer length:
\begin{equation*}
\mathcal L_{0,2}(z_1, z_2)= \frac{dz_1dz_2}{(z_1-z_2)^2} - \frac{d\tilde{x}_1d\tilde{x}_2}{(\tilde{x}_1-\tilde{x}_2)^2}
\end{equation*}
with $\tilde{x}_i$ being the value of the $\tilde{x}$-coordinate of the point of the curve \eqref{othercurve} defined by $z_i$.  Thus $\mathcal L_{0,2}$ is interpreted as the difference of the differentials $w_{0,2}$  on the curve and on the $\tilde{x}$-sphere.

It turns out that with $w_{0,2}$ replaced by $\mathcal L_{0,2}$ and with $w_{0,1}=\tilde{y}d\tilde{x}$ the recursion kernel $K$ defined by \eqref{K} coincides with our kernel \eqref{HOkernel} if $c^2=-2\epsilon$. Moreover, starting at $w_{g,n}$ with $2g-2+n\geq 2$, the recursion formula \eqref{recursion} only uses the difference 
$w_{0,2}(-z,z_k) - w_{0,2}(z,z_k)$, see discussion preceding Corollary \ref{corollary_wgn}. This difference is obviously unchanged if we replace $w_{0,2}$ by $\mathcal L_{0,2}$. Therefore, starting from the second generation of $w_{g,n}$, that is from the indices $(g,n)$ such that $2g-2+n\geq 2$, the recursion formula \eqref{recursion} applied to the harmonic oscillator curve and the recursion from \cite{Chapman} coincide.
 
It matters however  what we use as $w_{0,2}$ in the recursion formula \eqref{recursion} for the calculation of the first generation of $w_{g,n}$,  such that  $2g-2+n=1$, that is $w_{1,1}=\mathcal L_{1,1}$ and $w_{0,3}=\mathcal L_{0,3}$. These two differentials are not obtained in \cite{Chapman} by the topological recursion procedure on the curve \eqref{othercurve} but rather considered as initial values for the recursion. They are calculated by taking the Laplace transform of the numbers of the integral ribbon graphs of the corresponding types.

We note that we obtain all of the $\mathcal L_{g,n}=w^H_{g,n}$ with $2g-2+n>0$, including $w^H_{1,1}=\mathcal L_{1,1}$ and $w^H_{0,3}=\mathcal L_{0,3}\,,$ by the topological recursion on the harmonic oscillator curve  using the original definition of Eynard and Orantin \cite{EO}. This is the advantage of using the harmonic oscillator curve rather than the curve \eqref{othercurve}.

As is shown in \cite{MulasePenkava}, the functions $L_{g,n}$ as functions of $z_j$ after the change of variables \eqref{change} coincide with the Poincar\'e polynomials $P_{g,n}$ \eqref{Poincare1}, which implies \eqref{harmonicPoincare}. The next theorem shows that we recover the Poincar\'e polynomials
of $RG_{g,n}$ as integrals of a special form of the differentials $w^H_{g,n}$.

\begin{theorem}
\label{thm_integrals}
Let the pair $(g,n)$ be such that $2g-2+n>0$ and $w^H_{g,n}$ be the multi-differentials obtained by the topological recursion of Eynard-Orantin on the harmonic oscillator curve.  Then
we have  for the Poincar\'e polynomial $P_{g,n}$ \eqref{Poincare1} defined in \cite{MulasePenkava}:
\begin{equation*}
P_{g,n}(z_1,\dots, z_n) =\left(-\frac{c^2}{2\epsilon}\right)^{{2g-2+n}} \int_{-1}^{z_1}\dots \int_{-1}^{z_n} w^H_{g,n}(z'_1,\dots, z'_n).
\end{equation*}
\end{theorem}

\begin{remark}
Note that  $z=-1$ specifies  the point at infinity on the sheet of the curve $y^2=x^2-c^2$ which corresponds to $y=-\sqrt{x^2-c^2}$.
A similar integration of $w_{g,n}$ was used, for example, in \cite{BergereEynard}.
\end{remark}

{\it Proof of Theorem \ref{thm_integrals}.}
We will now prove that the polynomial $\Omega_{g,n}(z_1,z_2,\dots, z_n)$ defined by 
\begin{equation*}
\Omega_{g,n}(z_1,\dots, z_n) = \left(-\frac{c^2}{2\epsilon}\right)^{{2g-2+n}} \int_{-1}^{z_1}\dots \int_{-1}^{z_n} w^H_{g,n}(z'_1,\dots, z'_n)
\end{equation*}  
coincides with the Poincar\'e polynomial \eqref{Poincare1} defined  in  \cite{MulasePenkava}, where $2g-2+n > 0$. The recursion obeyed by  $P_{g,n}$ is presented in (1.6) of   \cite{MulasePenkava} (reproduced below in (\ref{poinc})) with  initial data $P_{1,1}(z_1)$ and $P_{0,3}(z_1,z_2,z_3)$.  Using expressions (\ref{wgn12}) for $w^H_{1,1}$ and $w^H_{0,3}$, we find that these initial data from \cite{MulasePenkava} agree with our $\Omega_{1,1}$ and $\Omega_{0,3}$:

\begin{eqnarray*}
\Omega_{1,1} (z_1) &=&  - \frac{(1+z_1)^4}{384 z_1^2} \left(z_1 - 4 + \frac{1}{z_1} \right) \, ,\nonumber \\
\Omega_{0,3}(z_1,z_2,z_3) &=&  - \frac{(1+z_1)(1+z_2)(1+z_3)}{16} \left(1 + \frac{1}{z_1 z_2 z_3} \right)     \, .
\end{eqnarray*}
We emphasize, however, that in contrast with  \cite{MulasePenkava} where these two polynomials are determined independently through ribbon graphs, here they follow from applying the topological recursion of Eynard-Orantin to the harmonic oscillator curve.

  We will now establish that the derivative $\partial_{z_1}\Omega_{g,n}$ obeys the recursion for $\partial_{z_1} P_{g,n}$ given in (5.1) of \cite{MulasePenkava}.
 We   integrate 
 both sides of the recursion (\ref{wgnrecur}) for $w^H_{g,n}$ over  $z_i$  from  $-1$ to $z_i$ for all $i=2,\dots, n$, multiply by $(-c^2/2 \epsilon)^{2g-2+n}$  and divide by $dz_1$ to obtain

\begin{multline}
\partial_{z_1} \Omega_{g,n}(z_1,z_2,\dots, z_n) = \frac{1}{32 }\frac{(1-z_1^2)^3}{z_1^2}  \partial_s\partial_t \Omega_{g-1,n+1}(s,t,z_2, \dots, z_n)\Big{|}_{s=t=z_1} 
\\
+ \frac{1}{32 }\frac{(1-z_1^2)^3}{z_1^2}  \sum_{\substack{g_1+g_2=g\\I\coprod J=\{2,\dots, n\}}}^{\mbox{stable } } \partial_{z_1} \Omega_{g_1, |I|+1}(z_1,z_I)\partial_{z_1}\Omega_{g_2, |J|+1}(z_1,z_J)
\\
+\frac{1}{16}\sum_{j=2}^n   \int_{-1}^{z_j}  ~D_{z_1,z'_j} \partial_{s} \Omega_{g,n-1}(s, z_{[n]_{1,j}}) \frac{ds}{dz_1}.
\label{bfr}
\end{multline}

As previously, we consider the  action of the integral of the operator $D$ \eqref{operd} on even powers of $s$:

\begin{multline}
  \int_{-1}^{z_j}  ~D_{z_1,z'_j} \frac{ds}{s^{2k}} \frac{1}{dz_1} =  \mathcal D_{z_1,z_j}  \frac{1}{s^{2k}} 
+ \frac{(1-z_1^2)^3}{z_1^6} \sum_{s=0}^{1-k} z_1^{2(1-k-s)} + \frac{(1-z_1^2)^3}{z_1^2} \sum_{t=0}^{k-3} \frac{1}{ z_1^{2(k-t)}} 
\\
\left. + \frac{1}{ z_1^2}\left(  \frac{1-3z_1^2 + 3z_1^4}{ z_1^4} + \frac{1-3z_1^2}{z_1^2} +1  \right)  \right.
\label{hsg}
\end{multline}
with $\mathcal D_{z_1,z_j}  \frac{1}{s^{2k}}$  given in (\ref{curlyd}).  The last three terms on the right hand side of (\ref{hsg}) 
can be combined to give

\begin{equation*}
  \int_{-1}^{z_j} ~D_{z_1,z'_j} \frac{ds}{s^{2k}}\frac{1}{dz_1} =  \frac{(1-z_1^2)^2}{z_1^{2k+2}}  + \mathcal D_{z_1,z_j}  \frac{1}{s^{2k}} 
.
\end{equation*}

In order to compare with (5.1) of \cite{MulasePenkava}, one extracts a factor $z_j/ (z_1^2-z_j^2)$ from  the term containing the operator $\mathcal D_{z_1,z_j}$,
giving

\begin{multline*}
 \int_{-1}^{z_j}  ~D_{z_1,z'_j} \frac{ds}{s^{2k}} \frac{1}{dz_1} 
=  \frac{(1-z_1^2)^2}{z_1^{2k+2}}  +  \frac{z_j}{(z_1^2-z_j^2)}  
\biggl\{ 
  (1-z_1^2)^3 
\left( \frac{1}{z_1^{2k+2}} - \frac{1}{z_j^{2k-4}z_1^6} \right)  \bigl(H(k-3)  + H(k-1) \bigr)   \\
 + \frac{(1-z_1^2)^3}{z_1^6 z_j^{2k-4}} - \frac{(1-z_j^2)^3}{z_j^{2k+2}}
 \biggr\}   \, ,
\end{multline*}
where $H(n)$ is the discrete Heaviside step function with the convention $H(n)=1$ for $ n \geq 0$ and zero otherwise.
This  can easily be  checked to simplify to the following expression, valid for all values of $k$:

\begin{equation*}
  \int_{-1}^{z_j}  ~D_{z_1,z'_j} \frac{ds}{s^{2k}} \frac{1}{dz_1}
=  \frac{(1-z_1^2)^2}{z_1^{2k+2}}  +  \frac{z_j}{(z_1^2-z_j^2)}  
\frac{(1-z_1^2)^3}{z_1^6}  
\left( \frac{1}{z_j^{2k}} - \frac{1}{z_1^{2k}} \right)  . 
\end{equation*}

Using this result in (\ref{bfr}), we obtain

\begin{multline*}
\partial_{z_1} \Omega_{g,n}(z_1,z_2,\dots, z_n) = \frac{1}{32}\frac{(1-z_1^2)^3}{z_1^2}  \partial_s\partial_t \Omega_{g-1,n+1}(s,t,z_2, \dots, z_n)\Big{|}_{s=t=z_1} 
\\
+ \frac{1}{32}\frac{(1-z_1^2)^3}{z_1^2}  \sum_{\substack{g_1+g_2=g\\I\coprod J=\{2,\dots, n\}}}^{\mbox{stable } } \partial_{z_1} \Omega_{g_1, |I|+1}(z_1,z_I)\partial_{z_1} \Omega_{g_2, |J|+1}(z_1,z_J)
\\
+ \frac{1}{16} \frac{(1-z_1^2)^2}{z_1^{2}}  \partial_{z_1} \Omega_{g,n-1} (z_{[n]_{j}})  +\frac{1}{16}\sum_{j=2}^n \frac{z_j}{(z_1^2-z_j^2) }  
\biggl\{  \frac{(1-z_1^2)^3}{z_1^2} \partial_{z_1} \Omega_{g,n-1}(z_{[n]_{j}}) -   \frac{(1-z_j^2)^3}{z_j^2}   
\partial_{\partial z_j}  \Omega_{g,n-1}(z_{[n]_{1}}) \biggr\}  ~ ,
\end{multline*}

which coincides with (5.1) of  \cite{MulasePenkava}. 

To obtain the recursion formula for the $\Omega_{g,n}$, we simply integrate over $z_1$ from $-1$ to $z_1$, trivially giving

\begin{multline}
 \Omega_{g,n}(z_1,z_2,\dots, z_n) = \int_{-1}^{z_1} \left(  \frac{1}{32}\frac{(1-{z'_1}^2)^3}{{z'_1}^2}  \partial_s\partial_t \Omega_{g-1,n+1}(s,t,z_2, \dots, z_n)\Big{|}_{s=t=z'_1} \right.
\\
+ \frac{1}{32}\frac{(1-{z'_1}^2)^3}{{z'_1}^2}  \sum_{\substack{g_1+g_2=g\\I\coprod J=\{2,\dots, n\}}}^{\mbox{stable } } \partial_{z'_1} \Omega_{g_1, |I|+1}(z'_1,z_I)\partial_{z'_1}P_{g_2, |J|+1}(z'_1,z_J)
+ \frac{1}{16} \frac{(1-{z'_1}^2)^2}{{z'_1}^{2}}  \partial_{z'_1} \Omega_{g,n-1} (z_{[n]_{j}})  
\\
+\left. \frac{1}{16}\sum_{j=2}^n \frac{z_j}{({z'_1}^2-z_j^2) }  
\biggl\{  \frac{(1-{z'_1}^2)^3}{{z'_1}^2} \partial_{ z'_1} \Omega_{g,n-1}(z_{[n]_{j}}) -   \frac{(1-z_j^2)^3}{z_j^2}   
\partial_{ z_j}  \Omega_{g,n-1}(z_{[n]_{1}}) \biggr\}  \right)dz'_1 ~ . \label{poinc}
\end{multline}
which agrees with (1.6) of \cite{MulasePenkava}.

We have thus shown that $ \Omega_{g,n} ( z_1 \ldots z_n) = P_{g,n}(z_1 \ldots z_n)$. 
$\Box$

\appendix
\markboth{}{}
\renewcommand{\thesection}{\Alph{section}}
\numberwithin{equation}{section}
\section{Appendix}
\section*{The residue of the unstable term}
\label{sect_appendix}

 The objective of this appendix is to prove the residue formulas used in Lemma \ref{lemma_residues}.  The residues of (\ref{res1}) are trivial to obtain. 
In order to prove (\ref{res2}), let us first define
\begin{equation*} F(a,k) :=  \sum_{n=0}^a \frac{n}{z_1^{2k-2n} z_j^{2n}} \,. \end{equation*}

In terms of this expression, one finds for the residue at $z=0$  
 \begin{multline*} \underset{z=0}{\rm Res}\, K^{H}(z,z_1)\left( - \frac{2z^2+2z_j^2}{(z^2-z_j^2)^2}  \right) \frac{(dz)^2}{z^{2k}}=  \frac{\epsilon dz_1}{8 c^2} \left( \frac{F(k+1,k)}{z_1^4 } +(1-3z_j^2)    \frac{F(k,k)}{z_1^2z_j^2}   \right. \\  \left. +(3z_j^2-3)  \frac{F(k-1,k)}{ z_j^{2}} +(3-z_j^2)z_1^2 \frac{F(k-2,k) }{ z_j^2} - z_1^4 \frac{F(k-3,k)}{ z_j^2} \right)~.\end{multline*}
 In each sum we introduce a new summation variable $p$ such that the sums all start at zero and such that the exponent of $z_j$ is  $-2p-2$. This gives 

\begin{equation*} 
- \frac{\epsilon dz_1}{8 c^2} \left(\sum_{p = 0}^{k} \frac{2 p +1 }{z_j^{2 p +2} z_1^{2k-2 p + 2 }} 
-3  \sum_{p=0}^{k-1} \frac{2p+1}{z_j^{2 p+2} z_1^{2k-2 p  }} 
%x
+ 3 \sum_{p= 0}^{k-2} \frac{2 p+1}{z_j^{2 p+2} z_1^{2k-2 p -2 }} 
- \sum_{p = 0}^{k-3} \frac{2 p+1}{z_j^{2 p+2} z_1^{2k-2 p -4 }}  \right) .
\end{equation*}
Next, we isolate  the sums up to $k-3$, obtaining

\begin{multline}
-\frac{\epsilon dz_1}{8 c^2} \left( \frac{(1-z_1^2)^3}{z_1^2 z_j^2} \sum_{p = 0}^{k-3} \frac{2p+1}{z_j^{2 p} z_1^{2k-2 p }}  
 + \frac{2k+1}{z_j^{2k+2} z_1^2}  H(k)
 %\\ 
 +  \frac{(2k-1)(1-3 z_1^2)}{z_j^{2k} z_1^{4}}  H(k-1)   \right.
 \\ \left.
 +
 \frac{(2k-3)(1-3z_1^2 +3 z_1^4)}{z_j^{2k-2} z_1^{6}}   H(k-2)  \right) \, .  \label{resinf} 
\end{multline}
 
Now let us  consider  the residue at infinity.
We define
\begin{equation*} G(a,k) := \sum_{n=0}^a \frac{n}{z_1^{2n+2k} z_j^{2-2n}} \,.\end{equation*}
The residue at infinity gives
\begin{multline*}
 \underset{z=\infty}{\rm Res}\,  K^{H}(z,z_1) \left( - \frac{2z^2+2z_j^2}{(z^2-z_j^2)^2}  \right) \frac{(dz)^2}{z^{2k}}=  \frac{\epsilon dz_1}{8 c^2} \left( 
z_1^4  \, G(2-k,k) + z_1^2 (z_j^2-3) \, G(1-k,k)  \right. \\ \left. +(3-3z_j^2)  \, G(-k,k)    + \frac{1}{z_1^2}  (3z_j^2-1)  \, G(-k-1,k)  - 
\frac{z_j^2}{z_1^4}   \, G(-k-2,k) \right) ~. 
\end{multline*}
 Again, we introduce a parameter $p$, this time such that the exponent of $z_j$ is always  $2 p$. This gives

\begin{multline*}  -  \frac{\epsilon dz_1}{8 c^2} \left(
-\sum_{p=0}^{1-k} (2 p+1) z_j^{2p} z_1^{2-2p-2k} 
+3 \sum_{p=0}^{-k} (2p+1) z_j^{2p} z_1^{-2p-2k} \right. \\ \left.
- 3 \sum_{p=0}^{-k-1} (2p+1) z_j^{2p} z_1^{-2-2p-2k}
+
\sum_{p=0}^{-k-2} (2p+1)  z_j^{2 p} z_1^{-4-2 p - 2 k} \right) ~.
\end{multline*}
We isolate the sums up to $1-k$ to obtain
   \begin{multline} - \frac{\epsilon dz_1}{8 c^2} \left(
   \frac{(1-z_1^2)^3}{z_1^4}   \sum_{p=0}^{1-k} \frac{ (2p+1)  z_j^{2 p}}{ z_1^{2 p + 2 k}} 
    + \frac{2k+1}{  z_j^{2k+2} z_1^{2}}  H(-1-k)  \right. \\ \left.
   + \frac{(2k-1)}{z_j^{2k} z_1^{4}} (1-3 z_1^2)   H(-k) 
   + \frac{(2k-3)(1-3z_1^2+3z_1^4)}{z_1^6 z_j^{2k-2}}    H(1-k) \right) ~. \label{reszero}
   \end{multline}
  To get the final result, we combine  (\ref{reszero}) and (\ref{resinf}). Note that   all the terms with 
  Heaviside functions have complementary conditions on the values of $k$ and they combine to give expressions valid for all values of $k$ and we obtain (\ref{res2}).

\end{document}